\documentclass[conference]{IEEEtran}

\usepackage{amssymb}
\usepackage{graphicx}
\usepackage{rotating}
\usepackage{booktabs}
\usepackage{multirow}
\usepackage[T1]{fontenc}
\usepackage{amsmath}
\usepackage[slide,vlined,linesnumberedhidden,ruled]{algorithm2e}
\usepackage{color}
\usepackage{url}

\usepackage{atbegshi}
\AtBeginDocument{\AtBeginShipoutNext{\AtBeginShipoutDiscard}}

\ifCLASSOPTIONcompsoc

\else

\fi

\ifCLASSINFOpdf

\else

\fi

\IEEEoverridecommandlockouts

\begin{document}

\title{On the Reverse Engineering of the Citadel Botnet}

\author{\IEEEauthorblockN{Ashkan Rahimian\IEEEauthorrefmark{2}\IEEEauthorrefmark{3}, Raha Ziarati\IEEEauthorrefmark{2}\IEEEauthorrefmark{3}, Stere Preda\IEEEauthorrefmark{2}\IEEEauthorrefmark{3}, and Mourad Debbabi\IEEEauthorrefmark{2}\IEEEauthorrefmark{3}}

\IEEEauthorblockA{
\IEEEauthorrefmark{2}~National Cyber-Forensics and Training Alliance Canada\\
\IEEEauthorrefmark{3}~Computer Security Laboratory, Concordia University\\
Montreal, Quebec, Canada\\
}}

\IEEEcompsocitemizethanks{\IEEEcompsocthanksitem This article is the final draft post-refereeing version of the paper appeared in FPS 2013. DOI: 10.1007/978-3-319-05302-8\_25}

\IEEEcompsoctitleabstractindextext{
\begin{abstract}
Citadel is an advanced information-stealing malware which targets
financial information. This malware poses a
real threat against the confidentiality and integrity of personal
and business data. A joint operation was recently conducted by the FBI
and the Microsoft Digital Crimes Unit in order to take down Citadel
command-and-control servers. The operation caused some disruption in
the botnet but has not stopped it completely. Due to the complex
structure and advanced anti-reverse engineering techniques, the
Citadel malware analysis process is both challenging and time-consuming.
This allows cyber criminals to carry on with their attacks while
the analysis is still in progress. In this paper, we present the
results of the Citadel reverse engineering and provide additional
insight into the functionality, inner workings, and open source
components of the malware.
In order to accelerate the reverse
engineering process, we propose a clone-based analysis methodology.
Citadel is an offspring of a previously analyzed malware called Zeus;
thus, using the former as a reference, we can measure and quantify
the similarities and differences of the new variant. Two types of
code analysis techniques are provided in the methodology, namely
assembly to source code matching and binary clone detection.
The methodology can help reduce the number of functions requiring 
manual analysis.
The analysis
results prove that the approach is promising in Citadel malware
analysis. Furthermore, the same approach is applicable to similar malware
analysis scenarios.
\end{abstract}

}

\maketitle

\IEEEdisplaynotcompsoctitleabstractindextext

\section{Introduction}

Making headlines in recent months \textit{(March 2013 - July 2013)} has been an offspring of Zeus malware known as Citadel. Cyber criminals behind the Citadel malware have stolen over 500 million dollars from online bank accounts~\cite{cit500}. Zeus is a prolific Trojan that has been stealing information since 2007. In 2011, its source code was leaked on the internet and became available to the underground community. Since then, several malware have been developed based on the Zeus source code. Citadel has been employed by botnet operators to steal banking credentials and personal information~\cite{iwc13,del12}. In addition, Citadel has features that extend beyond targeting financial institutions. Spying capabilities, such as video capture, is an example of such features that literally enables cyber criminals to collect anything from a victim's machine. The malware also acts as ransomware and scareware in order to extort money from victims. Reverse engineering is often the primary step taken to gain an in-depth understanding of a piece of malware; however, it is both a challenging and time-consuming process which requires a great deal of manual intervention.

The major objectives of this paper are to reverse engineer the Citadel malware and gain more insight into its structure and functionality. In particular, the objectives can be summarized as follows:
\begin{enumerate}
    \item Quantify the similarities and differences between Citadel and Zeus malware.
    \item Obtain additional insight into online open source components used in Citadel.
    \item Accelerate the reverse engineering process of similar malware variants.
\end{enumerate}
To enhance and speed up the process, a new approach termed clone-based analysis is employed in this study. This paper illustrates the usefulness of the proposed approach in the analysis of new variants of a malware family. In this scenario, a preceding malware \textit{P} is presumed to be analyzed and understood. If a variant \textit{V} uses portions of the \textit{P} code, the approach will highlight the shared portions. Consequently, disregarding the clones could reduce the analysis time.
In a more general case where \textit{P} is unknown, the approach can still provide insight into the components of \textit{V} in comparison to other sources.

The main contributions of this paper are three-fold: first, a detailed reverse engineering analysis of the Citadel malware is presented and its functionality is described. Second, a new methodology for reverse engineering malware is proposed. This methodology significantly decreases malware analysts' efforts and reduces the analysis time.
Third, the similarity between the Citadel malware and the Zeus malware is precisely quantified. Additional insights are also provided into the open-source components used in the Citadel malware.

This case study was chosen for a number of reasons. First, Citadel and Zeus are real threats against confidentiality, integrity and availability of information systems. Cyber criminals are constantly enhancing their tools for gaining access to personal and financial data. The profitability of such crimeware tools in the underground market depends on the timeliness and support for new vulnerabilities. For this reason, malicious developers often reuse all or parts of existing components during their incremental development process. As a result, it is quite probable that they leave fingerprints of previously analyzed malcode on new releases. Clone-based analysis serves as a beneficial tool in such situations due to its potential for producing quick results. Integrating a clone-based analysis in the reverse engineering process will significantly reduce the overall analysis time. The second benefit of this case study is that it allows us to leverage our developed tools such as \textit{RE-Source} \cite{fps12} and \textit{RE-Clone} \cite{cff12} in reverse engineering sophisticated malware. The lessons learned during the analysis would bring new opportunities for future extensions of our tools. Third, the analysis provides us with practical solutions for mitigating future threats in a timely fashion. Once the analysis is performed on Zeus and Citadel, new Zeus-based malware variants with shared components can be more quickly analyzed.

The remainder of this paper is organized as follows. Section~\ref{sec:method} is dedicated to explaining our methodology in studying the malware. Section~\ref{sec:dynamic} details the dynamic analysis and explains the debugging process and memory forensic approaches. The main features of the Citadel malware are also described in this section. Section~\ref{sec:static} presents the static analysis and the steps which led to the actual de-obfuscated code. Section~\ref{sec:Clone-based Analysis} presents the clone-based analysis. The threat mitigation is briefly presented in Section~\ref{sec:mitigate}, and the conclusion is drawn in Section~\ref{sec:conclusion}.

\section{Methodology}
\label{sec:method}

\begin{figure}[tbp]
    \centering
        \includegraphics[width=0.9\linewidth]{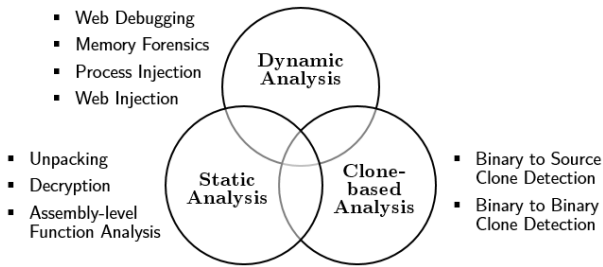}
				
    \caption{The overlap in reverse engineering methodologies.}
    \label{fig:venn}
\end{figure}

Static and dynamic analysis are commonly used in studying malware \cite{pma12,imc09}. Static analysis
focuses on malware code, inspecting its structure and functionality without execution.
In contrast, dynamic analysis deals with behavior monitoring during the malware execution.
In general, the process of malware reverse engineering is a combination of these two approaches, which is both time-consuming and costly. The success of these approaches is tightly coupled with the functionalities of the tools and skills of the reverse engineer \cite{mfg12,ida11}.

To enhance and accelerate the process in analyzing the Citadel malware, another dimension is considered in our study as shown in Figure~\ref{fig:venn}. This new dimension is called clone-based analysis. In brief, the clone-based analysis identifies the pieces of code in Citadel malware that originate from other malware and open-source applications. This step is performed automatically by leveraging the tools that are designed and developed in our security lab \cite{fps12,cff12}. There are two main advantages in considering this extra dimension in the static analysis: first, it avoids dealing with low-level assembly code in situations where the corresponding high-level code is available; second, it prevents reverse engineering parts of the malware that have already been analyzed. This approach is very promising, especially in scenarios similar to Citadel that share a significant portion of code with a previously reverse engineered malware like Zeus~\cite{pst10}. The process of assembly to source code matching is performed using the \textit{RE-Source} framework~\cite{fps12}, and the binary clone matching is carried out using \textit{RE-Clone}~\cite{cff12}.

The proposed methodology is composed of three processes. Each process comprises several steps. We elaborate each step in the following sections:

\begin{enumerate}
    \item \textit{Static Analysis Process}
        \begin{itemize}
            \item The disassembly is reviewed for finding obfuscated segments, decoder stubs, and embedded file images. The feasibility of static data decryption is assessed. Switching to dynamic analysis for code and data decryption may be necessary.
            \item A suitable circumvention strategy is adopted for bypassing the anti-static protection of the malware. Having a de-obfuscated/decrypted disassembly is a prerequisite for the clone-based analysis process.
            \item Control flow analysis and data flow analysis are applied to gain an understanding of the crypto algorithms and encoding/decoding functionality.
        \end{itemize}

    \item \textit{Dynamic Analysis Process}
        \begin{itemize}
            \item A debugging environment is set to execute the binary, attach the debugger, set the breakpoints, control the unpacking, dump the process memory, generate an executable image, and save the process to file. The dumping process is repeated according to the analysis scenario.
            \item System calls are monitored, malware activities are logged, network traffic is captured, downloaded files are backed up, and the communication protocol is observed. In addition, the interesting artifacts are extracted.
        \end{itemize}

    \item \textit{Clone-based Analysis Process}
        \begin{itemize}
            \item Using the unpacked and de-obfuscated disassembly, a search is performed for standard and open-source components by applying the assembly to source-code matching technique of \textit{RE-Source}~\cite{fps12}. The data matching technique encompasses two threads of \textit{online} and \textit{offline} analysis. This step is repeated for all process memory dumps and the set of matched projects are stored as \textit{online} analysis results.
            \item An \textit{offline} analysis is performed for assigning the functionality tags according to API call classification in \textit{RE-Source}. Function labels are updated, the proportion of assembly functions in each functionality group is calculated, and the functionality tags are reviewed based on the scenario.
            \item Using the unpacked and de-obfuscated disassembly, a binary clone matching is performed against the previously analyzed malware binaries in \textit{RE-Clone}. The occurrences of \textit{inexact} and \textit{exact} clones are then recorded.
            \item The outputs of assembly to source-code matching and binary clone matching are combined for quantifying the similarities and difference of malware variants. The results draw a high-level picture of the code.
            \item The clones are selectively used to guide the static and dynamic analyses. In order to speed up the process, the clones are removed and the focus of analysis is shifted to the original (non-clone) functions.
        \end{itemize}
\end{enumerate}

As illustrated in Figure~\ref{fig:venn}, three connected processes are defined in the proposed methodology. In the Citadel case study, the dynamic analysis focuses on web debugging, memory forensics, process injection and web injects. An important aspect in this process is the observation of malware's behavior in response to controlled inputs. On the other hand, the static analysis process focuses on assembly-level functions. De-obfuscation could occur in the overlapping area of these two methods. Unpacking and decryption are relevant examples that fall in this area. It is assumed that a database of previously analyzed code is available during the analysis. Code search engines provide an interface to online open source code repositories. Similarly, an offline code repository is maintained for storing the malware assembly code and the results of previous analysis sessions. One advantage of the clone-based analysis is that it can guide the dynamic and static steps: it highlights the important directions that the other two processes should follow by eliminating code clones, recognizing library functions, and providing additional comments. The analysis focus is thus shifted to non-clone parts of the payload, resulting in a shorter analysis time frame.

\section{Dynamic Analysis}
\label{sec:dynamic}

The purpose of the dynamic analysis process is to execute the malware and monitor its behavior in a controlled environment. Many tools and techniques are available for debugging malware \cite{ghp09,mfg12}. Sandboxing is a common technique in dynamic analysis used for running untrusted code in a virtual setting; however, modern malware are well-equipped with anti-virtual machine protection against popular tools such as \textit{Oracle VirtualBox} and \textit{VMWare Workstation}. The malware can easily sense whether it is running on a virtual machine by checking certain artifacts in memory or on disk. As a result, the malware might change its normal behavior by taking an alternative execution path, thereby hindering the analysis. Malware can go even one step further and try to exploit the virtual machine vulnerabilities in order to gain access to the host operating system. Successful dynamic analysis may thus require caution and pre-processing steps. Debugging Citadel is challenging due to the built-in anti-debugging and injection capabilities, but the protection can be circumvented by choosing the right strategy. As it will be discussed in Section \ref{sec:Clone-based Analysis}, \textit{RE-Source} can provide informative tags such as {\ttfamily ADB}, {\ttfamily PSJ} or {\ttfamily AVM} for functions that potentially contain anti-debugging, process injection, or anti-virtual machine functionality. Upon the first execution, Citadel begins the infection process based on an embedded attack configuration.
\subsection{Citadel Infection Process}

The Citadel bot operates in several modes. Upon the first
execution, the dropper is in the \textit{installation mode}. First,
it unpacks and decrypts itself into memory. It then creates a
copy of the binary file and stores it in the {\ttfamily\%AppData\%}
folder under a randomly generated sub-folder and file name. The bot
file is referred to as \textit{Random.exe} in this context. As an
example, the output path could be similar to:
\textit{...\textbackslash AppData\textbackslash
Roaming\textbackslash Random\textbackslash Random.exe.} The bot also
generates a batch file for removing the installation code. Checking for
the existence of this path is a way to determine whether the system
has been infected by the malware. Once the \textit{Random.exe} is
run from the new location, a sequence of similar steps are taken for
unpacking and decrypting the bot. Subsequently, the bot switches to
 \textit{injection mode} and injects itself into the Explorer process and
its child processes.

\begin{figure}
    \centering
        \includegraphics[width=0.85\linewidth]{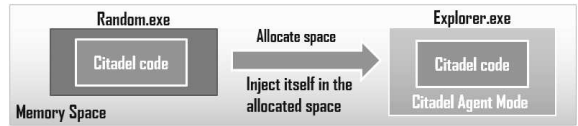}
    \caption{Citadel process injection and agent mode.}
    \label{fig:injectCit}
\end{figure}

The injection step is dependent upon the
privileges of the user who runs the bot as well as the version of the
operating system. Following the injection, the bot process is
terminated and the installation files are removed. The bot also
updates the registry and adds an entity so that it will execute each
time the operating system boots up. The registry path would appear
as: \textit{HKU\textbackslash...\textbackslash
Software\textbackslash Microsoft\textbackslash Windows\textbackslash
CurrentVersion\textbackslash Run\textbackslash Rad.} The
\textit{Random.exe} is identical to the dropper except for
the flag bytes located at the end of the file. This portion is
encrypted and is used for controlling the bot mode. Thus, although the 
two executables are very similar, their behavior is
different as the \textit{Random.exe} operates only in \textit{agent and
injection modes}. Upon each system startup, the bot initiates the intelligence gathering process
as demonstrated in Figure~\ref{fig:injectCit}.

\subsection{Debugging and Memory Forensics}

After setting up the analysis environment and infecting it with the malware, the bot execution can be monitored and controlled using a scriptable debugger~\cite{imm13,ida13}. Several techniques are available for hiding the debugger process from the bot and gaining more control over the debugger~\cite{ghp09}. A web debugger or a network protocol analyzer is used for monitoring the HTTP network communication of the malware. Citadel encrypts the command-and-control (C\&C) network traffic with RC4; therefore, the crypto keys are required to intercept the commands and view the stolen data.
One way to find the keys is through debugging and setting hardware breakpoints on functions that precede network communication.

As will be discussed in Section~\ref{sec:Clone-based Analysis}, such network-related functionality can be identified through the {\ttfamily NET}, {\ttfamily WNT} and {\ttfamily CRY} tags assigned in the offline analysis. 
\begin{figure}
    \centering
        \includegraphics[width=0.9\linewidth]{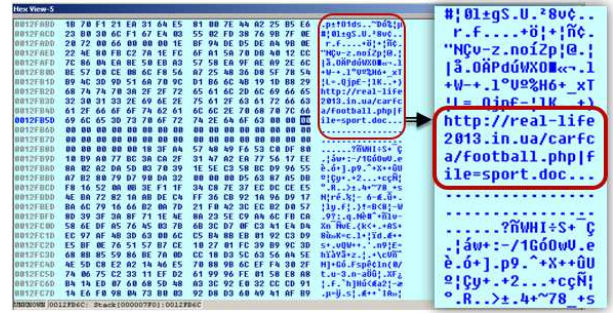}
    \caption{Decoded Citadel config file name and location.}
    \label{fig:configCit}
\end{figure}
Upon successful installation, the bot checks for Internet
connectivity and tries to connect to embedded C\&C addresses in
order to announce its availability. The bot sends requests such as
{\ttfamily POST /carfca/basket.php HTTP/1.1} or {\ttfamily POST
/carfca/file.php HTTP/1.1} to the server. The server then replies
and sends the encrypted config file to the bot. One major difference
between Zeus and Citadel is in the way they handle the transmission
of the configuration file. It was possible to find the location of
the Zeus config file and download it with minimal effort, whereas in
Citadel, it is more difficult to obtain the config file during the
analysis. Citadel uses dynamic APIs and decrypts strings in
memory during execution. This can be considered as an extra
layer of protection that prevents the
config file from being easily detected. Figure~\ref{fig:configCit}
shows one of the decrypted links to a Citadel C\&C server which
hosts the  encrypted ``\textit{sport.doc}'' config file. During the
debug, the bot allocates memory for new segments and overwrites the
memory space with decrypted code and data. The
zero values in Figure~\ref{fig:configCit} show the bytes that are
yet to be overwritten by data. Blocking the malware's access to the
requested C\&C and modifying its timing mechanism will force the
malware to enumerate the list of alternative embedded C\&C servers.

\begin{figure}[htb]
    \centering
        \includegraphics[width=1.0\linewidth]{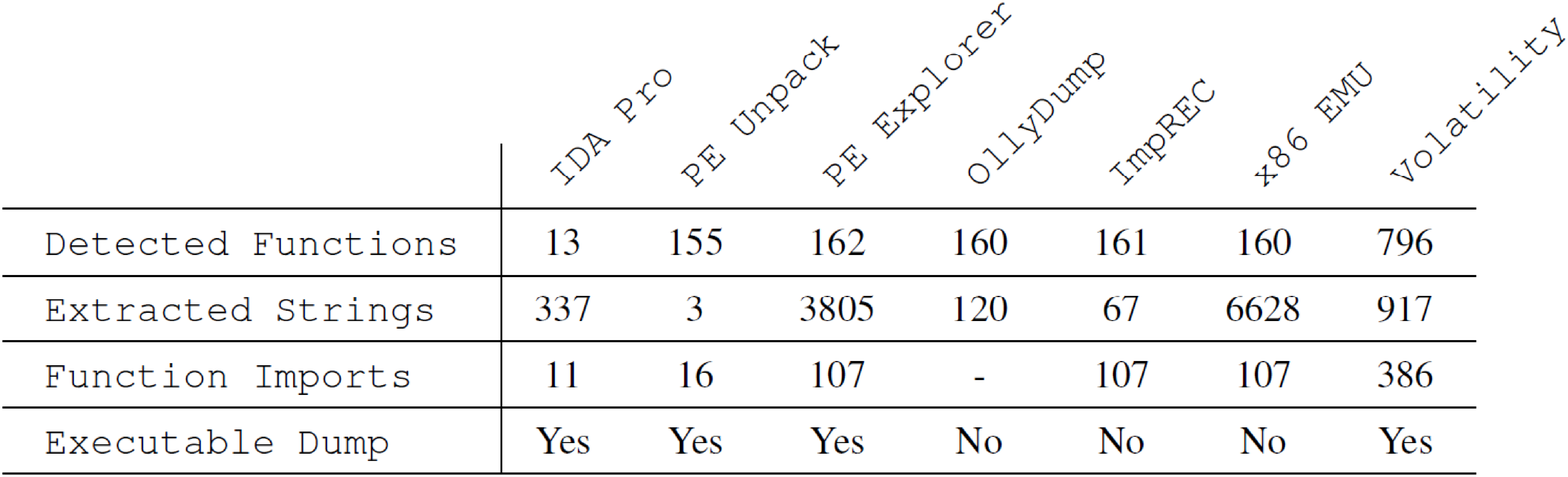}
    \caption{Unpacking stages of Citadel binary.}
    \label{fig:unpackDeobuscate}
\end{figure}

Several tools and plug-ins are available for dumping memory,
reconstructing import tables, and fixing PE headers.
\textit{OllyDump} and \textit{ImpRec} are examples of such tools for
unpacking Citadel \cite{pma12,mfg12}.
\textit{Volatility}~\cite{vol13} was the most versatile and
straightforward tool for memory forensics that was used in this
project. It automatically builds the import tables and generates the
executable versions of the unpacked binary. \textit{Volatility} was
utilized for creating executable process dumps and retrieving decrypted
strings from memory. Figure~\ref{fig:unpackDeobuscate} lists
the utilized tools and shows the number of detected functions, extracted strings, and identified
function imports during different stages of the unpacking process.

\subsection{Citadel Attack Configuration}

\begin{figure}[h]
    \centering
        \includegraphics[width=0.95\linewidth]{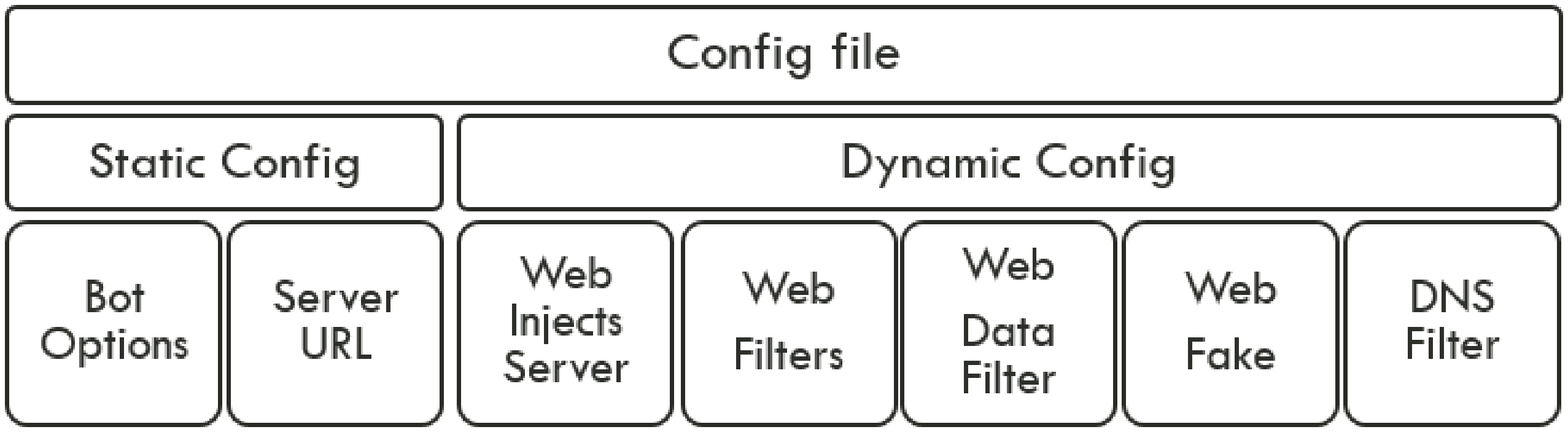}
    \caption{Structure of Citadel configuration file.}
    \label{fig:configInternal}
\end{figure}

The bot options are set in the configuration file. This file
contains two sections for static and dynamic configuration as
depicted in Figure~\ref{fig:configInternal}. The bot builder reads
this file and embeds the settings in the generated \textit{bot.exe}.
The bot encryption key is also defined in this file. The
static config section is where the options for the initial attack
are set. The setting for web injects are defined in the
dynamic config section. Web injects are used for tricking users
into revealing confidential information such as additional passwords
or PINs. Since the man-in-the-middle attack (DLL hooking) occurs in libraries such as \textit{wininet.dll} or
\textit{nspr4.dll}, the victim user might not be able to distinguish
the injected data from the genuine page. The result of injection
could be in the form of extra fields, text boxes or warning messages.
In comparison to Zeus, Citadel has a few extra features such as the
anti-virus and security software evasion mechanism. In addition, the DNS
filter enables the bot to block the victim from accessing
security-related websites and downloading new updates and patches.
Consequently, the machine is made more vulnerable to future
attacks. A DNS redirection technique is used for implementing this
feature. The config file includes a list of blocked
websites and the corresponding redirected IP addresses. The report
in \cite{del12} provides a list of Citadel DNS filter domains. This
type of DNS poisoning attack does not modify the Windows
\textit{Hosts} file. The settings related to the dynamic
configuration can be updated by the C\&C server according to
predefined rules set by the botmaster. For instance, new modules can
be remotely installed for country-specific web inject which target
online banking accounts, automatic money transfer, and
ransom~\cite{tdc13,del12}. The encrypted configuration file can be
obtained by capturing the bot traffic and replaying a crafted
request in debugging.

\section{Static Analysis}
\label{sec:static}

In this section, we describe the main steps taken during the static analysis of Citadel. The static malware analysis process normally starts by disassembling the malware binary. However, the initial disassembled code may not draw a complete picture of the original code due to different layers of obfuscation. Disassembling the Citadel malware using \textit{IDA Pro} \cite{ida13} results in a packed binary containing merely 13 functions, 11 imports, and 337 strings. The binary was compressed, encrypted, and employed anti-reverse engineering techniques. Our static analysis therefore began by de-obfuscating the malware. According to the first process of the proposed methodology, static and dynamic techniques should be interleaved for advancing the analysis.

\subsection{Unpacking Step}

Not surprisingly, the malware was packed with a non-standard packing scheme, thus automatic unpacking tools such as \textit{UPX} could not be used and manual unpacking was necessary. To unpack the malware, a combination of static and dynamic techniques was used. The packed binary was executed in \textit{Immunity debugger} \cite{imm13} until the unpacking stub decompressed the binary in memory. Once the unpacking procedure was completed, the unpacking stub transferred the execution to the original entry point of the binary by making a jump from one segment to another. At this moment, \textit{Volatility} \cite{vol13} was used to dump the unpacked version of the binary's process out of memory and generate an executable unpacked version of the binary. The generated binary contained roughly 800 functions, 386 imports, and more than 900 decrypted strings.

\subsection{Code Decryption Step}

\begin{figure}[hbp]
    \centering
        \includegraphics[width=0.7\linewidth]{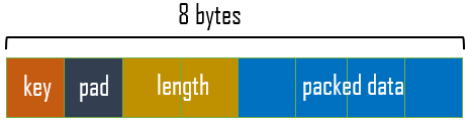}
    \caption{Structure of the encrypted data.}
    \label{fig:pad}
\end{figure}

The unpacking allowed the static analysis to be resumed. After this step, there still remained encrypted portions in the binary code. One of the interesting points was located at the address of {\ttfamily 0x0040336} in our sample. An in-depth examination of the function which cross-referenced this portion revealed the structure of encrypted data and the decryption mechanism.  As shown in Figure~\ref{fig:pad}, the structure size is 8 bytes and consists of 4 chunks. The key for string decryption is embedded in the binary file.
Algorithm~\ref{alg:strDec} presents the decryption procedure used for decrypting the data. It
helped us recover more than 300 strings and 45 C\&C commands from the packed data in the binary.

\begin{algorithm}[hbp]\caption{String Decryption Procedure}\label{alg:strDec}
\tcc{The command for decrypting embedded strings}
{\For{${j}$~in~$range$ length}  { 
~ 

    UNPACKED\_DATA = \texttt{join}(char(PACKED\_DATA[\emph{j}]) \^{} \emph{j} \^{} key)
    }
}

\end{algorithm}

\subsection{Crypto Algorithms}

\begin{figure}[tbp]
    \centering
        \includegraphics[width=1.00\linewidth]{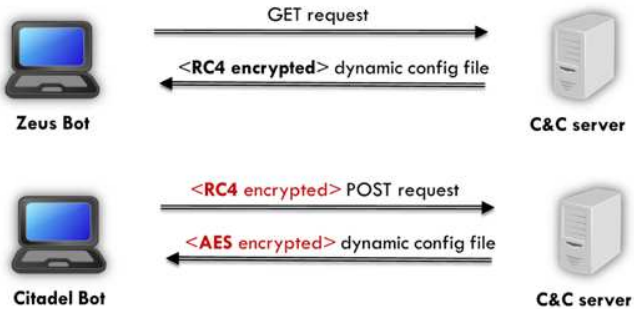}
    \caption{Communication messages for retrieving the configuration file.}
    \label{fig:comm2}
\end{figure}

\begin{figure}[htbp]
    \centering
        \includegraphics[width=1.00\linewidth]{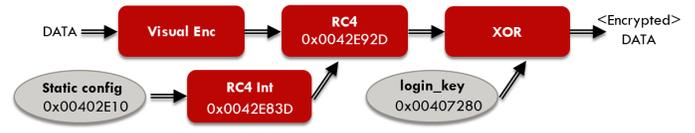}
    \caption{Citadel RC4 encryption process.}
    \label{fig:RC4}
\end{figure}

\begin{figure}[htbp]
    \centering
        \includegraphics[width=1.00\linewidth]{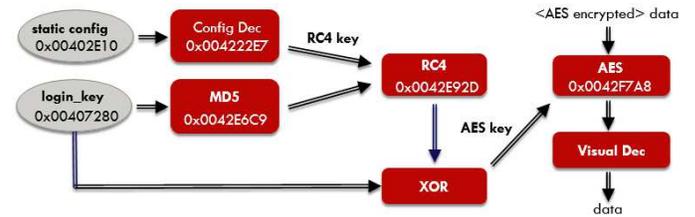}
    \caption{Citadel AES decryption process.}
    \label{fig:AES}
\end{figure}

The Zeus malware suffered from two main weaknesses: receiving an RC4-encrypted configuration file from a C\&C server in
response to a plain {\ttfamily GET} request, and reusing
nonrandom values for encrypted messages. To overcome these weaknesses,
significant improvements concerning crypto
algorithms have been made in Citadel. As shown in
Figure~\ref{fig:comm2}, the Citadel C\&C server requires
a specially crafted RC4-encrypted {\ttfamily POST} message to send
the configuration file. In addition, in order to provide 
better security, the configuration file is encrypted using AES.
The Citadel authors have used a composition of different
ciphers as shown in Figures~\ref{fig:RC4} and \ref{fig:AES}. The RC4 encryption
(Figure~\ref{fig:RC4}) starts by a customized encoding (obfuscation) mechanism known as
\textit{Visual Encrypt (VE)}.
\begin{algorithm}[htbp]\caption{Visual Encrypt Procedure}\label{alg:visEnc}
{void \texttt{Crypt::VisualEncrypt}(void
*buffer, DWORD size) \{

\For{ (DWORD i = 1; i < size; i++)}  { 
~ 

((LPBYTE)buffer)[i] \^{} = ((LPBYTE)buffer)[i - 1];
    }~\}
}
\end{algorithm}

The input to the algorithm is an encoded
buffer. The \textit{VE} code is provided in
Algorithm~\ref{alg:visEnc}. This function was used in Zeus for
crypto purposes as well. After the XOR operation, the non-standard
RC4 initialization routine generates a 0x100-byte key based on the
static configuration data embedded in the binary. The output of the
routine is a new RC4 key that is used in RC4 encryption function
along with the customized XOR-ed data. Finally, performing an XOR on
the RC4 output and the login key embedded in the binary results in
the RC4 encrypted data. Given \textit{login\_key=lkey} and
\textit{VE=encode}, the functionality can be stated as:
\textit{out} = \textit{lkey} XOR
RC4$_{\text{rkey}}$\textit{(encode(in))}. Therefore,
\textit{out}=Enc(\textit{in}).
The AES decryption is depicted in
Figure~\ref{fig:AES}. The configuration decryption routine takes
as input the embedded static configuration data and produces as output the RC4 key. 
The MD5 hashed login key and the embedded RC4 key are fed to the RC4
routine. Subsequently, the AES key is generated by performing an XOR on the
output of the RC4 routine and the login key.
This key is used by the AES decryption function.
Finally, the \textit{Visual Decrypt (VD)}
function (Algorithm~\ref{alg:visDec}) takes the result of the AES routine and decodes the
decrypted data. The process can be formulated as:
AES$_{\text{key}}$ = MD5(\textit{lkey}) XOR RC4$_{\text{rkey}}$. Given
\textit{VD=decode}, the output can be stated as:
\textit{out = decode(}AES$_{\text{AES\_key}}$(\textit{in})). The
weakest point in the crypto process is that it
is based on static config data, which shows that the authors lack competency in security algorithms and cipher composition.

\begin{algorithm}[htbp]\caption{Visual Decrypt Procedure}\label{alg:visDec}
{void \texttt{Crypt::VisualDecrypt}(void
*buffer, DWORD size) \{

\If{ size > 0  } {

\For{ (DWORD i = size $-$ 1; i > 0; i$--$)}  {
~

 ((LPBYTE)buffer)[i] \^{}= ((LPBYTE)buffer)[i - 1];
    }
}

\}

 }
\end{algorithm}

\section{Clone-based Analysis}
\label{sec:Clone-based Analysis}

The third process of the proposed methodology focuses on clone-based analysis, which can be applied to complement the process of reverse engineering. It could be particularly helpful in reducing the time required for the static analysis phase. In this context, two techniques are taken into account for quantifying the similarities between Citadel and Zeus samples. The first approach uses \textit{RE-Source} in order to reveal the open-source building blocks of the malware. The second approach utilizes
\textit{RE-Clone} for binary code matching. The major steps in the clone-based methodology can be enumerated as follows: (1) identification of standard algorithms and open-source library code in the malware disassembly, (2) assigning meaningful labels to assembly-level functions based on API classification, (3) commenting the assembly code based on a predefined dictionary of malware functions, and (4) applying a window-based search and comparison mechanism for finding the pre-analyzed code components.

\subsection{Assembly to Open-Source Code Matching}

\begin{figure}[tbp]
    \centering
        \includegraphics[width=1.00\linewidth]{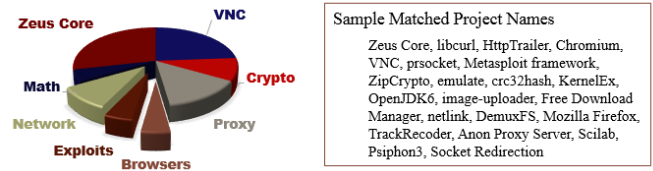}
    \caption{Matched features with open-source projects.}
    \label{fig:opensource}
\end{figure}

\begin{figure*}[htbp]
    \centering
        \includegraphics[width=0.8\linewidth]{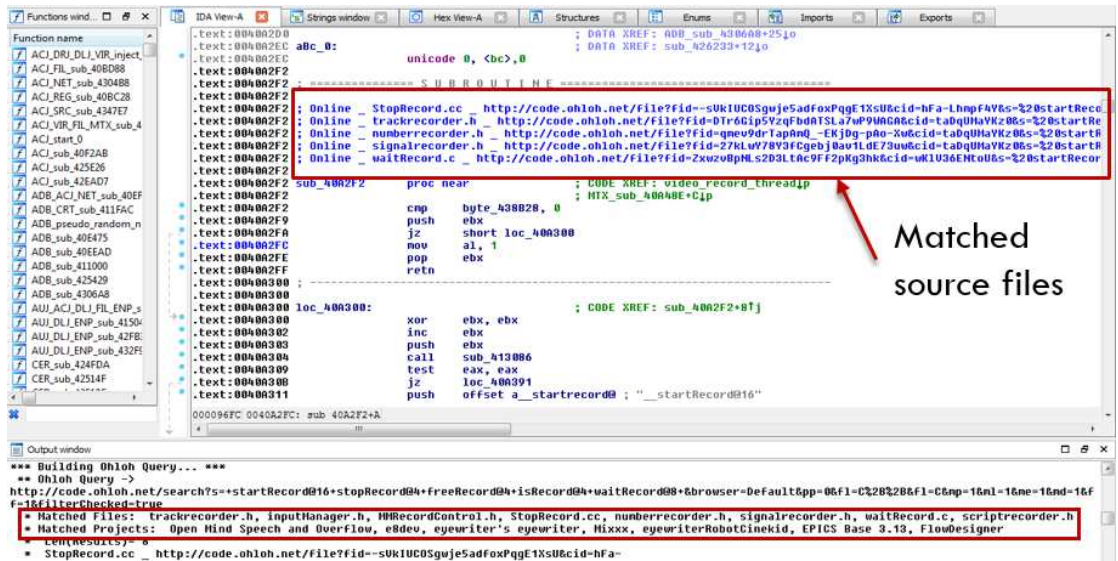}
    \caption{The output of RE-Source pointing to video capture source code.}
    \label{fig:IDA}
\end{figure*}

The \textit{RE-Source} framework \cite{fps12} has been used for
extracting assembly-level features from Citadel. 
This framework examines assembly functions in both online and offline phases in order to find source files that share features with the disassembly.
The key steps of the framework are: (1) extraction of
interesting features, (2) feature-based query encoding, (3) query
refinement for online code search engines, (4) request/response
processing, (5) data extraction and parsing, (6) reporting results
and updating comments, and (7) feature-based offline analysis.
Different features are considered for online and offline analysis.
During the online analysis phase, \textit{RE-Source} revealed the
correlation between function-level features of Citadel and several
open-source projects.  The video capture capability of the malware
was unleashed through links to source files such as:
\textit{MHRecordContol.h}, \textit{stopRecord.c},
\textit{trackerRecorder.h}, \textit{signalRecorder.h},
\textit{waitRecord.c}, etc. (See Figure~\ref{fig:IDA}). This observation was
further supported by occurrences of strings such as
\textit{``\_startRecord\@16''} during the dynamic API
de-obfuscation. Moreover, a ``{\ttfamily video\_start}'' C\&C
command was also found in this process. Even though screen
capture is a common feature in modern malware, live video capture
capability is a new feature which is only seen in complex and progressive samples.
It should be noted that the online analysis results of \textit{RE-Source} suggesting video-related capability were the outcome of an approximate code matching process. Although the matching process was not perfect, it was sufficiently accurate to reveal the functionality context in this case. Similarly, \textit{RE-Source} had commented the code with references to other open-source projects such as those listed in Figure~\ref{fig:opensource}. The number of matched projects in each category determines the size of each pie slice.
Many Zeus-based malware variants have appeared online since the
release of Zeus source code in 2011. Having access to Zeus source
code enabled us to match Citadel binary against Zeus source code.
The pie chart in Figure~\ref{fig:opensource} shows the general
categories of open-source projects used in Citadel. Apart
from the detached slices, Citadel and Zeus share a considerable
amount of code related to the core, VNC, crypto, and proxy
functionalities. However, the differences can be summarized in network
communication code, new exploits and browser-specific code for web
injects.

\subsection{Offline Analysis and Functionality Tags}
\textit{RE-Source} can also be used for tagging assembly functions
based on API calls and classifying functions according to their
potential functionality. When applied to the unpacked version of
Citadel, 652 functionality tags were detected by the offline
analyzer. A function is assigned several tags if it contains more
than one system call. Accurate functionality tags could convey
meaningful hints to the reverse engineer during the static analysis
phase. In conjunction with the code and data cross-referencing,
functionality tags can enrich the disassembly by highlighting the
final system calls in a multi-level function call hierarchy. Since
system calls serve as interaction points with the operating system,
having a high-level view of them could draw a more organized view of
the code.

\begin{figure}[htbp]
    \centering
        \includegraphics[width=1.0\linewidth]{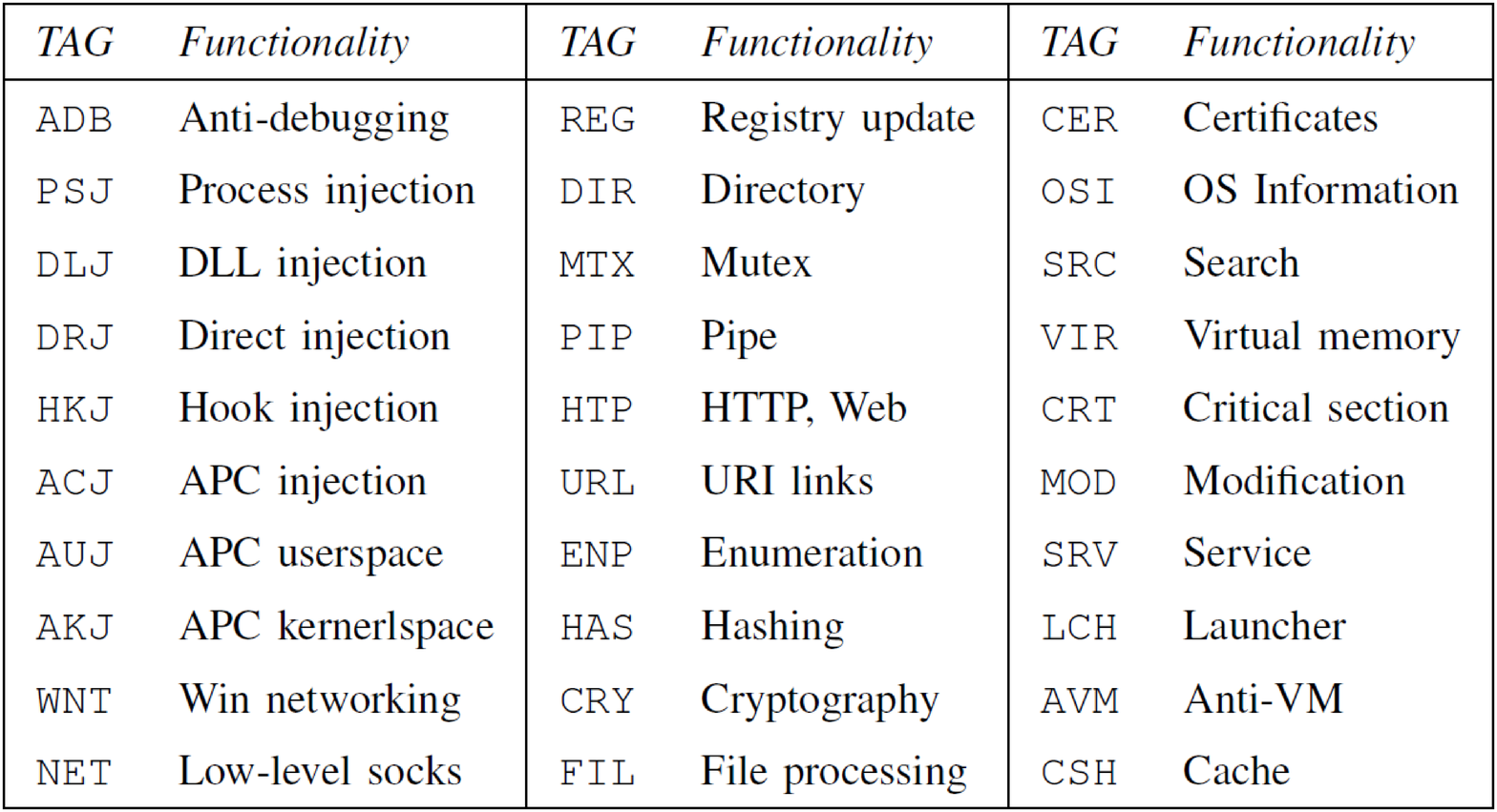}
    \caption{Functionality tags for offline analysis.}
    \label{fig:OtherFunctionalityTags}
\end{figure}

Functionality tags are not limited to simple system calls for
file processing or registry modifications. They can be composed of
several operations related to common malware behavior. 
New patterns can be defined for highlighting common
malicious code in downloaders, launchers, reverse shells, remote
calls and keyloggers based on the combination of several simple
system operations.
In this context, process memory modification and
code injection points are of great interest to the reverse engineer.
\textit{RE-Source} includes tagging categories such as
\textit{process injection}, \textit{launcher}, \textit{DLL
injection}, \textit{process replacement}, \textit{hook injection},
\textit{APC injection} and \textit{resource segment manipulation} in the offline analysis. 
Figure~\ref{fig:OtherFunctionalityTags} lists some of the available
functionality tags in the prototype.
A practical application of functionality tags is in disassembly comparison/synchronization of two malware variants. Instead of comparing the files by address, the code can be analyzed offline and the generated tags can be used as association criteria/sync points. In this process, the functions are sorted based on the assigned tags, and those with similar tags are analyzed side by side. This technique was specifically helpful in synchronizing the disassembly of Citadel versus Zeus.
\begin{figure}[tbp]
    \centering
        \includegraphics[width=0.65\linewidth]{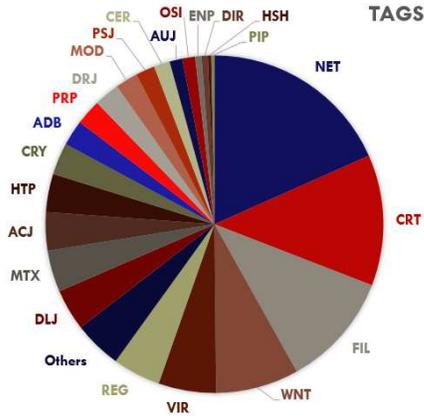}
    \caption{Functionality tags assigned by offline analysis.}
    \label{fig:tags}
\end{figure}
Figure~\ref{fig:tags} depicts  the detected functionality tags. The
pie chart sectors are proportional to the number of assembly
functions categorized under the same functionality group.
The {\ttfamily NET} tag was assigned to 60 functions related to
low-level socks.
Furthermore, 41 functions were tagged with
{\ttfamily CRT} (critical section objects) for mutual exclusion
synchronization. Similarly, 36 {\ttfamily FIL} tags were assigned to
file manipulating functions. The other tags, such as crypto, hashing,
search and code injection, were also identified during the analysis.
The {\ttfamily CRY} (crypto) and {\ttfamily HSH} (hashing) tags
provided an easy way of disassembly synchronization between Citadel
and Zeus as the slight differences between the assembly files had no
effect on the overall functionality group.

\begin{figure}[htbp]
    \centering
        \includegraphics[width=1.0\linewidth]{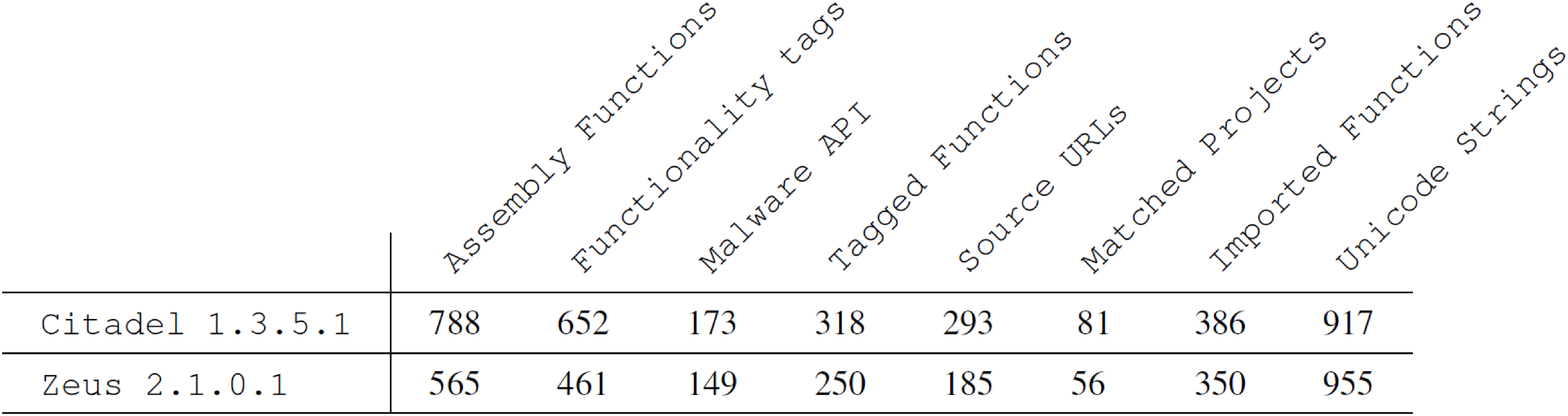}
    \caption{RE-Source analysis results.}
    \label{fig:RESourceAnalysisResults}
\end{figure}

Translated into quantifiable  terms,
Figure~\ref{fig:RESourceAnalysisResults} shows the output of
\textit{RE-Source} for Citadel vs. Zeus comparison. The numbers are
reported in accordance with occurrence of certain features such as
the number of assembly functions, API and functionality tags, common
API in malware, number of matched opens source components, imported
function calls, and Unicode strings. The results imply that the framework
has been successful in revealing the internal components of the malware.
The final outcome of assembly to source code matching is a list of source files alongside the description from the malware dictionary. This information provides valuable
insight into the potential functionality of the malicious code and facilitates the analysis.

\subsection{Binary Clone Analysis}

The malware analysis process can be accelerated by identifying and
removing the previously analyzed code fragments. The aim of binary
clone analysis is to compare the assembly file of a new binary
sample with a repository of analyzed code.
The result of this analysis is a set
of matched \emph{clones}. In this context, we rely on the
\textit{RE-Clone} binary clone detector tool~\cite{cff12} that
implements an improved version of the clone detector framework
proposed in~\cite{Saebjornsen}.
\textit{RE-Clone} considers the same problem
definition, that is the \emph{exact} and \emph{inexact} clone
detection, as stated in~\cite{Saebjornsen}.
 Exact clones share the same assembly features, i.e.,
mnemonics, operands and registers. The only difference is in memory
addresses. Inexact clones can be regarded as equal up to a certain
level of abstraction, which means that the number of common features must be greater
than a certain threshold.

\begin{figure}[htbp]
    \centering
        \includegraphics[width=1.00\linewidth]{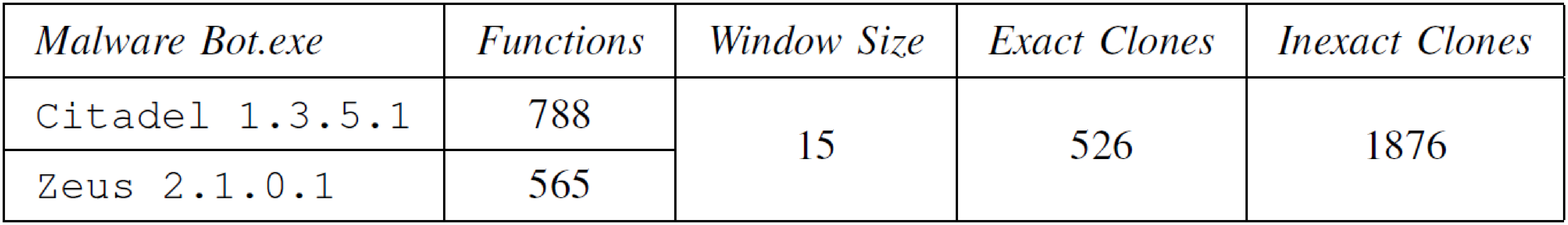}
    \caption{Binary clone detection results.}
    \label{fig:BinaryCloneDetectorResults}
\end{figure}

\begin{figure}[tbp]
    \centering
        \includegraphics[width=1.00\linewidth]{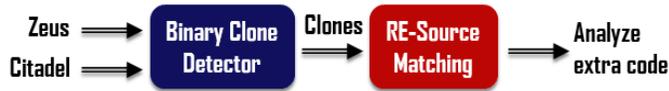}
    \caption{Code analysis after clone elimination.}
    \label{fig:diff}
\end{figure}

The analysis parameters such as search
window size, normalization level and detection algorithm play a
significant role in the analysis results. These parameters are set
according to each analysis scenario. After marking the detected code
fragments as clones, the focus of analysis is shifted to non-analyzed
and new code segments.
The core components of the Zeus malware has been thoroughly studied
in~\cite{pst10}. The source and binary files are also available
online. A new Zeus variant can therefore be compared against
the existing files in order to measure the similarity and detect the
potential exact and inexact clones. This analysis is also applicable
to finding the additional functions of the new malware variant.
Figure~\ref{fig:BinaryCloneDetectorResults} shows the results of the
binary clone matching process. The samples
share 526 exact binary clones with a window size of 15 instructions.
In other words, almost 93\% of Zeus assembly code also appears in
Citadel. These clones form approximately 67\% of the Citadel binary.
This analysis highlights the remaining 33\% of the Citadel
assembly to be analyzed. Thus, a significant amount of time is
saved by disregarding the clones. \textit{RE-Clone} shows the 
address of each clone in the disassembly. Furthermore,
the remaining functions can be examined in \textit{RE-Source}
before the manual analysis process is begun by the reverse engineer.
This approach is depicted in Figure~\ref{fig:diff}. The 1876 inexact
clones reported by the tool include multiple combinations of regions
that also contain the exact clones.

\begin{figure*}[htbp]
    \centering
        \includegraphics[width=0.9\linewidth]{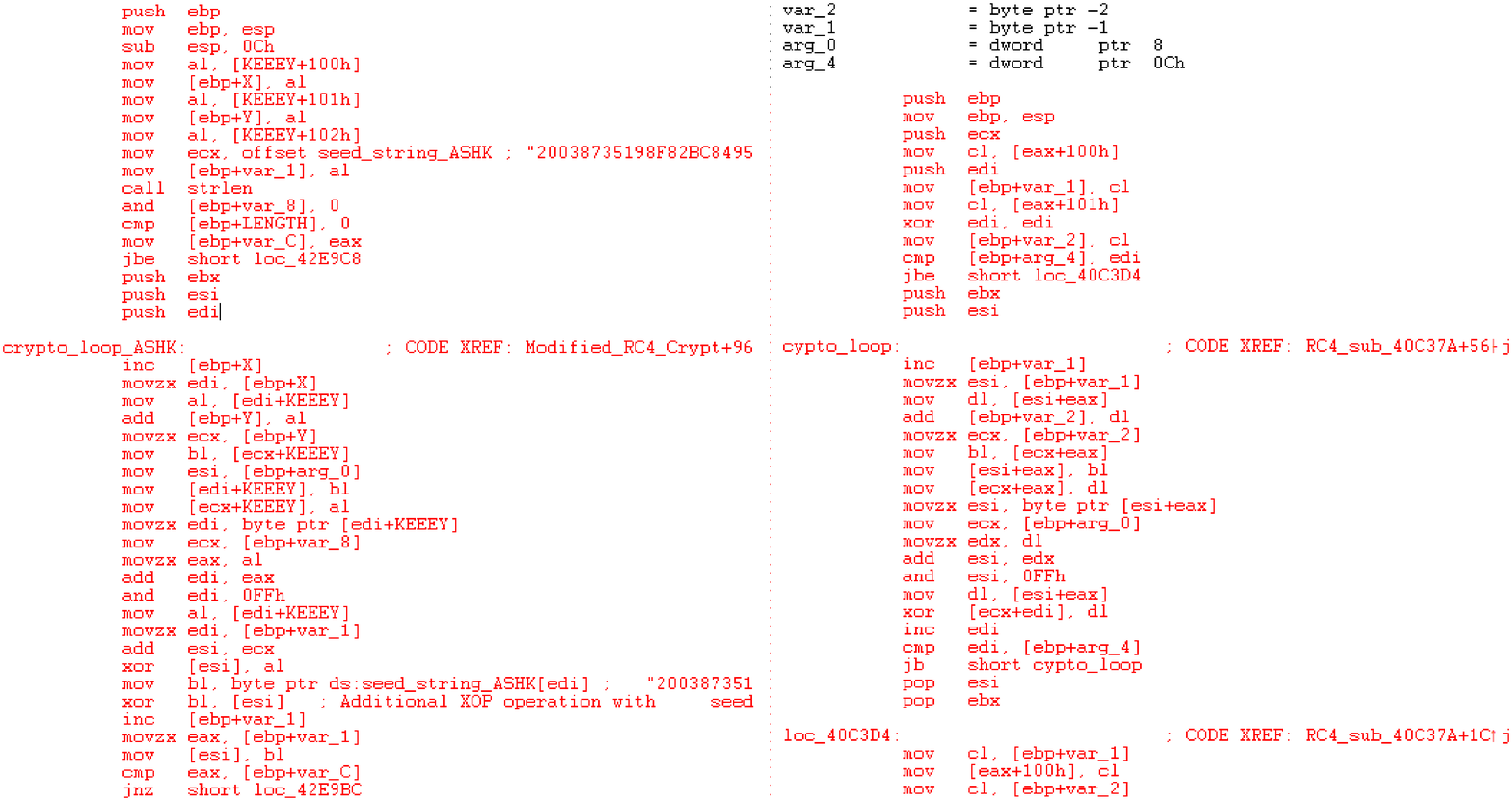}
    \caption{An inexact clone detected in the RC4 encryption function (Citadel vs. Zeus).}
    \label{fig:RC4clone}
\end{figure*}

An interesting example of  crypt-related clones is the detection of
an inexact clone in the RC4 function that is used for encrypting
the C\&C network traffic. There are a few extra assembly
instructions in the Citadel version of the RC4 function. This clone
was found with a threshold of 0.8 and a two-combination inexact
clone search method. In this approach, each two-combination of
features is considered as a cluster. If more than 80\% of regions
appear in the same clusters, they are treated as inexact
clones. The red segments in Figure~\ref{fig:RC4clone} highlight the detected inexact clones.

\section{Threat Mitigation by Sinkholing}
\label{sec:mitigate}

In June 2013, the Microsoft Digital Crimes Unit reported on an
operation known as \emph{Operation b54} in collaboration with the FBI
to shut down Citadel C\&C servers~\cite{msd13}. As a result of this
operation, 1400 Citadel botnets around the world were interrupted
and redirected to sinkhole servers controlled by Microsoft. A
comprehensive list of the domain names is available in \cite{lst13}.
Although the operation significantly disrupted Citadel botnets and reduced the threat levels, it also affected honeypot systems
that were used for identifying and locating the malware creators
and distributors. Despite the overall success of the threat countermeasure, cyber criminals can still operate by infecting
new machines and controlling their bots using alternative servers.

\section{Related Work}

AnhLab~\cite{ahn12} presented a comprehensive static analysis of
Citadel malware. To the authors' knowledge, this report is the most
complete analysis on Citadel malware that has been released so far.
The process of infection, the structure of the malware binary, and
the malware's main functionalities and features are explained in detail
in this technical report. The report gives valuable insight into the
malware and its capabilities; however, the methodology and steps
that were taken for reaching the outcomes were not discussed. Furthermore,
despite mention in the report that Citadel is remarkably
similar to Zeus, the precise quantification of their similarity is
not provided. Only approximate resemblance percentages is given
without any details. To compare our analysis to this work, we
provided a new methodology for reverse engineering malware by
adopting clone-based analysis. Following our methodology, we
concisely explained the steps we took in reverse engineering Citadel
and the insight that we obtained through our study.

SophosLabs~\cite{cck12} provided a brief report on Citadel
malware. The major enhancements which occurred in Citadel compared to
Zeus is briefly explained in high-level in this report. No
explanation was provided about the process of reverse engineering
the malware and how the authors gained those insights. Indeed, this
report gave a modest overview about the Citadel malware without
digging into the details. CERT Polska~\cite{tdc13} also provided a
technical report on Citadel malware. Similar to the previously
mentioned report, this report was high-level and reviewed the
main features of Citadel without providing details. The reports
mainly provided statistics focusing on the impact of the malware and
its geographical distribution. The statistics were gathered based on
the traffic to the sinkhole server after the domain had been taken
down.

By leveraging the tools developed in our security lab, we quantified
the similarity between the Zeus and Citadel malware. These
results could be further refined by integrating other existing
techniques designed to automate malware analysis. For instance,
our binary clone detector could be extended with a CFG-like dimension.
For this purpose, we could benefit from the model proposed in~\cite{bonfante} 
which aims to identify the common code fragments between two executable files and
analyze the CFG subgraphs containing these fragments.

\section{Conclusion}
\label{sec:conclusion}

The Citadel malware targets confidential data and financial
transactions. It is an emerging threat against online privacy
and security. Citadel reverse engineering is challenging as it is
equipped with anti-reverse engineering techniques which hinder the
malware analysis process. As the number of incidents entailing new
malware attacks are increasing, agile approaches are required for
obtaining the analysis results in a timely fashion. The malware
reverse engineering process consists of two major stages of static
and dynamic analysis. This process can be accelerated and enhanced
by adding a new dimension for clone analysis. Instead of initiating
the process from scratch, a quick clone-based analysis can
easily highlight the similarities and differences between two
samples of the same family. The focus of analysis is then shifted to
the differing sections. We have presented a methodology along with
the tools and techniques for analyzing the Citadel malware. We have also compared Citadel with its predecessor, Zeus. The similarities
have been quantified as the result of two code matching techniques,
namely assembly to source and binary code matching. The same
methodology can be applied to other malware samples for providing
insight into the potential malware functionality. The results of
the malware analysis process can be added to a local code repository
and used as a reference for measuring the similarities between
future samples. They can also be used for improving the accuracy of
the results. Overall, the successful completion of our objectives
has led to underlining the best practices for supporting real-world
malware analysis scenarios.

\section*{Acknowledgment}
The authors would like to thank ESET Canada for their collaboration
and acknowledge the support of Mr. Pierre-Marc Bureau
and the guidance provided by Mr. Marc-Etienne Leveille on de-obfuscation.

\end{document}